\documentclass[runningheads]{llncs}
\usepackage[T1]{fontenc}
\usepackage{amsmath}

\usepackage{graphicx}
\usepackage{array}
\usepackage{amssymb}

\usepackage{hyperref}
\hypersetup{
    colorlinks=true,
    citecolor=blue,
    linkcolor=red,
    filecolor=magenta,      
    urlcolor=cyan,
    }
\usepackage{color}

\begin{document}
\title{Dynamic U-Net: Adaptively Calibrate Features for Abdominal Multi-organ Segmentation}
\titlerunning{Dynamic U-Net for Abdominal Multi-organ Segmentation}
%
\author{Jin Yang\inst{1} \and
Daniel S. Marcus\inst{1} \and
Aristeidis Sotiras\inst{1,2}}
\authorrunning{Jin Yang et al.}
%
\institute{Department of Radiology, Washington University School of Medicine, St. Louis, MO, USA \and
Institute for Informatics, Data Science and Biostatistics, Washington University School of Medicine, St. Louis, MO, USA \\
\email{yang.jin@wustl.edu}}

\maketitle              
\begin{abstract}
U-Net has been widely used for segmenting abdominal organs, achieving promising performance. However, when it is used for multi-organ segmentation, first, it may be limited in exploiting global long-range contextual information due to the implementation of standard convolutions. Second, the use of spatial-wise downsampling (e.g., max pooling or strided convolutions) in the encoding path may lead to the loss of deformable or discriminative details. Third, features upsampled from the higher level are concatenated with those that persevered via skip connections. However, repeated downsampling and upsampling operations lead to misalignments between them and their concatenation degrades segmentation performance. To address these limitations, we propose Dynamically Calibrated Convolution (DCC), Dynamically Calibrated Downsampling (DCD), and Dynamically Calibrated Upsampling (DCU) modules, respectively. The DCC module can utilize global inter-dependencies between spatial and channel features to calibrate these features adaptively. The DCD module enables networks to adaptively preserve deformable or discriminative features during downsampling. The DCU module can dynamically align and calibrate upsampled features to eliminate misalignments before concatenations. We integrated the proposed modules into a standard U-Net, resulting in a new architecture, termed Dynamic U-Net. This architectural design enables U-Net to dynamically adjust features for different organs. We evaluated Dynamic U-Net in two abdominal multi-organ segmentation benchmarks. Dynamic U-Net achieved statistically improved segmentation accuracy compared with standard U-Net. Our code is available at https://github.com/sotiraslab/DynamicUNet.
\end{abstract}
\section{Introduction}
Segmentation of multiple abdominal organs in medical images plays a crucial role in supporting clinical workflows, including diagnostic interventions and treatment planning \cite{gibson2018automatic}. Manual segmentation is labor-intensive, motivating the development of automatic tools. Specifically, U-Net and its variants have been applied to various automatic medical image segmentation tasks with great success \cite{oktay2018attention,ronneberger2015u,zhou2018unet++}. However, most of them focused on specific single organs, such as the liver \cite{kushnure2021ms,li2018h}, pancreas \cite{man2019deep,yang2023abdominal}, and kidney \cite{rani2022kub}. Multi-organ segmentation is still challenging due to the morphological complexity of the structures, the large inter-subject variations, and the large inter-organ variations in shape and size.

Among various multi-organ segmentation models, nnU-Net is a promising solution \cite{isensee2021nnu}. However, it relies on U-Net as the foundational segmentation architecture, which may pose three limitations in its architectural design, potentially hindering its ability to address challenges in multi-organ segmentation. First, its inherent limitation lies in the inability of convolution layers to adaptively calibrate features using global contextual information. Although attention modules can utilize global information \cite{oktay2018attention,sinha2020multi}, these are designed as independent modules to improve the whole network rather than improving convolution layers. The second limitation originates from applying downsampling layers (e.g., max-pooling and average-pooling) in the encoding path \cite{conze2021abdominal,ronneberger2015u}. These layers pool features equally by sliding a local window. However, some features are more discriminative than neighbors within each window. Thus, the applications of these downsampling layers cannot preserve discriminative details of organs \cite{gao2019lip,saeedan2018detail}. Although strided convolution is used for downsampling in nnU-Net \cite{isensee2021nnu} and other multi-organ segmentation models \cite{gibson2018automatic,zhang2020block}, it uses a fixed-sized convolutional window and cannot model the deformation of input features. Thus, these models fail to model variations in organ shape adaptively. Third, almost all multi-organ segmentation networks fuse upsampled features and features preserved via skip connections to facilitate information propagation \cite{gibson2018automatic,isensee2021nnu,roth2018multi,wang2019abdominal,zhang2020block}. However, repeated downsampling and upsampling lead to inaccurate correspondences (i.e., feature misalignment and organ mismatch) between them \cite{huang2021fapn,li2023dynamask}. Thus, the direct concatenation between them harms feature learning, resulting in the degradation of segmentation accuracy.

To address the aforementioned limitations, we propose Dynamically Calibrated Convolution (DCC), Dynamically Calibrated Downsampling (DCD), and Dynamically Calibrated Upsampling (DCU) modules. The DCC module can model long-range inter-dependencies and adaptively calibrate spatial-wise and channel-wise local features by global contextual information. The other two modules are designed based on (modulated) deformable convolutions \cite{dai2017deformable,zhu2019deformable}. Deformable convolutions have the capability of modeling the deformation and irregular shape of objects. The DCD module can handle the deformation and discrimination of features during downsampling. It dynamically determines the sampling interval scheme rather than adapting a fixed one in strided convolutions. Additionally, the DCD module can preserve deformation and discrimination, since it downsamples feature maps based on a higher information level adaptively rather than pooling features by a fixed scheme based on prior knowledge (e.g., max-pooling and average-pooling). Third, the DCU module is designed to adaptively align and calibrate upsampled features from lower resolution to their reference features from skip connections. It is achieved by adjusting each sampling location in a deformable convolutional kernel with an offset dynamically learned from the input feature maps. We integrated the above three modules into a standard U-Net to derive a novel Dynamic U-Net architecture for multi-organ segmentation. The integration of our three models enhances the capabilities of Dynamic U-Net to model organ deformations and effectively handle large variations among different organs. This adaptability holds promise for accurately segmenting organs with high variability, thereby improving the overall segmentation accuracy. 

Our contributions can be summarized as follows: (i) A \textbf{Dynamically Calibrated Convolution} module was designed to dynamically model long-range spatial-wise and channel-wise inter-dependencies. It exploits these interactions to adaptively calibrate local features based on global contextual information. (ii) A \textbf{Dynamically Calibrated Downsampling} module was designed to calibrate features adaptively by preserving the deformable and discriminative details during downsampling. This functionality enables the module to effectively handle deformations and large variations among abdominal organs. (iii) A \textbf{Dynamically Calibrated Upsampling} module was proposed to align and adaptively calibrate upsampled features with their skip-connected reference features. This maintains the consistency of deformation across different organs. (iv) We proposed the \textbf{Dynamic U-Net} for multi-organ segmentation to effectively handle the large variations among different organs and their complex anatomical structures. This is achieved by capturing and adaptively calibrating spatial and channel-wise deformable features, thereby improving segmentation performance.

\section{Methodology}
\subsection{Overall architecture}
\begin{figure}
\centering
\includegraphics[width=0.9\textwidth]{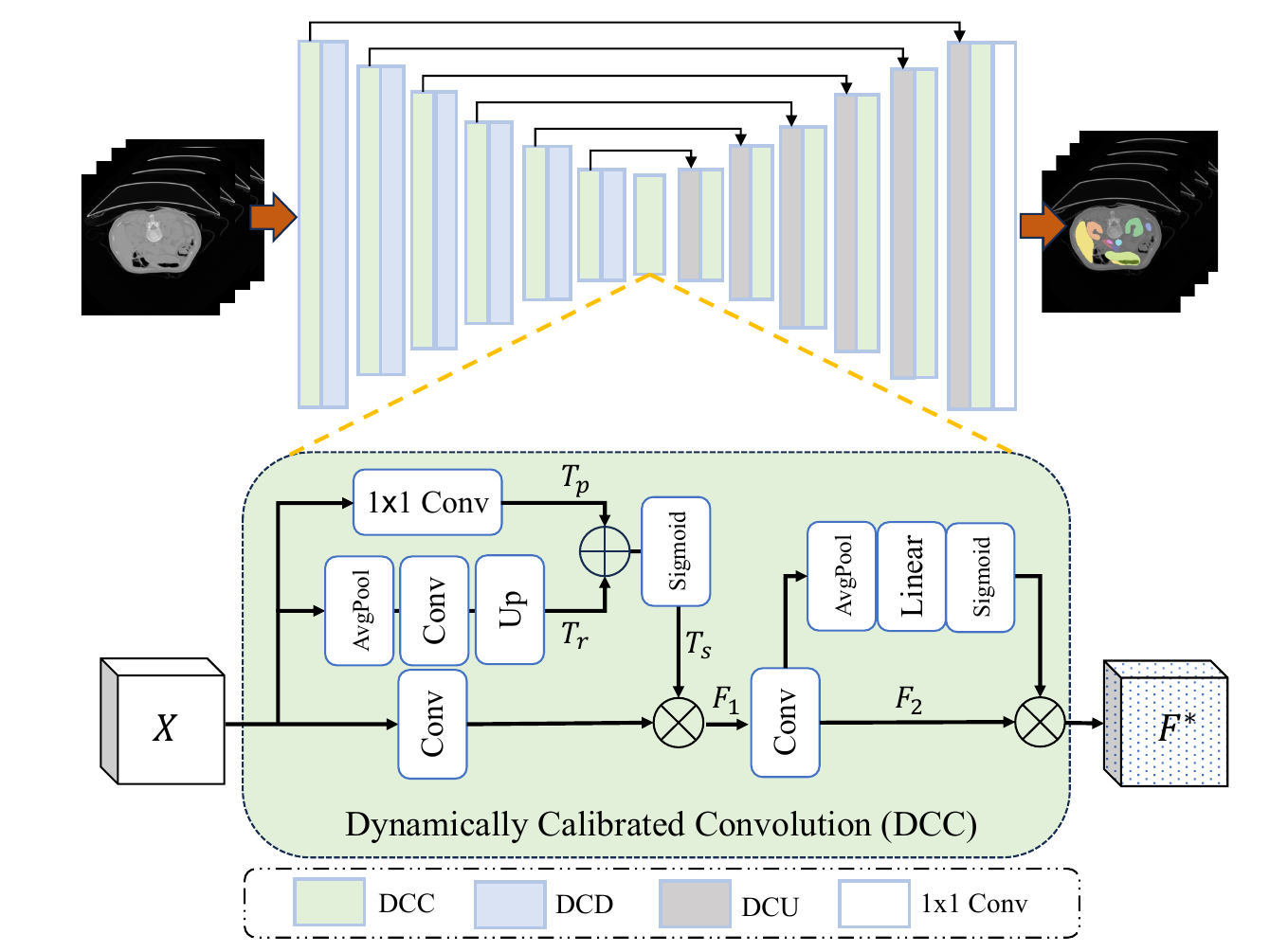}
\caption{The overall architecture of the Dynamic U-Net as well as the detailed architecture of the Dynamically Calibrated Convolution (DCC) module.} 
\label{fig:fig1}
\end{figure}

We proposed the Dynamic U-Net for segmenting multiple organs in abdominal CT images (Figure~\ref{fig:fig1}). This approach is based on incorporating the proposed DCC, DCD, and DCU modules into a standard 7-layer U-Net. Specifically, we used DCC modules to replace standard convolutions at each layer. We replaced the max-pooling layers with the DCD module and integrated the DCU module into the decoding path. The number of input feature maps was 1, and the numbers of feature maps in each layer were 32, 64, 128, 256, 512, and 512, respectively, with 512 feature maps in the bottleneck.
\subsection{Dynamically Calibrated Convolution (DCC) module}
To equip convolutions with the ability to adaptively calibrate spatial and channel features, we propose the Dynamically Calibrated Convolution (DCC) module (Figure~\ref{fig:fig1}). Two parallel paths are adopted to generate spatial-wise calibration information. The first path employs a $1\times1$ convolution layer $\mathcal{F}_{conv}^{(1)}(\cdot)$ to extract pixel-wise calibration information $\boldsymbol{T}_p\in \mathbb{R}^{C_{out}\times H \times W}$ from input feature maps $\boldsymbol{X}=[\boldsymbol{x}_1,\boldsymbol{x}_2,...,\boldsymbol{x}_{C_{in}}]\in \mathbb{R}^{C_{in}\times H \times W}$
\begin{equation}
    \nonumber
    \boldsymbol{T}_p = \mathcal{F}_{conv}^{(1)}(\boldsymbol{X}).
\end{equation}
However, this pixel-wise information ignores the utilization of neighboring contextual information. This is taken into account by the second path. This path utilizes average pooling $\mathcal{F}_{AvgPool}(\cdot)$ operation with filter $4\times4$ and stride 4 to capture and aggregate neighboring contextual information from input feature maps $\boldsymbol{X}\in \mathbb{R}^{C_{in}\times H \times W}$. Subsequently, a $3\times 3$ convolution $\mathcal{F}_{conv}(\cdot)$ operation is applied to generate region-wise calibration information $\boldsymbol{T}_r\in \mathbb{R}^{C_{out}\times \frac{H}{4} \times \frac{W}{4}}$. Lastly, nearest neighbor interpolation $\mathcal{F}_{up}(\cdot)$ is used to upsample the information to the original input dimension $\boldsymbol{T}_r\in \mathbb{R}^{C_{out}\times H \times W}$
\begin{equation}
\nonumber
    \boldsymbol{T}_r = \mathcal{F}_{up}(\mathcal{F}_{conv}(\mathcal{F}_{AvgPool}(\boldsymbol{X}))).
\end{equation}
In the next step, the pixel-wise information and neighboring region-wise information are fused, and a Sigmoid function $\mathcal{F}_{sig}(\cdot)$ is used to generate the spatial-wise calibration information $\boldsymbol{T}_s\in \mathbb{R}^{C_{out}\times H \times W}$
\begin{equation}
\nonumber
    \boldsymbol{T}_s=\mathcal{F}_{sig}(\boldsymbol{T}_p \oplus \boldsymbol{T}_r).
\end{equation}
Subsequently, this spatial-wise calibration information is used to calibrate local features via element-wise multiplication to generate calibrated feature maps $\boldsymbol{F}_1\in \mathbb{R}^{C_{out}\times H \times W}$. Here, local features are extracted by a $3\times 3$ convolution $\mathcal{F}_{conv}(\cdot)$ from input feature maps $\boldsymbol{X}\in \mathbb{R}^{C_{in}\times H \times W}$
\begin{equation}
\nonumber
    \boldsymbol{F}_1 = \mathcal{F}_{conv}(\boldsymbol{X})\otimes \boldsymbol{T}_s.
\end{equation}
In the subsequent step, another convolution layer $\mathcal{F}_{conv}(\cdot)$ is utilized to extract organ-specific spatial features and generate feature maps $\boldsymbol{F}_2\in \mathbb{R}^{C_{out}\times H \times W}$:
\begin{equation}
\nonumber
    \boldsymbol{F}_2 = \mathcal{F}_{conv}(\boldsymbol{F}_1).
\end{equation}
Finally, channel-wise calibration information is extracted by cascading an adaptive average pooling $\mathcal{F}_{AvgPool}(\cdot)$, a linear layer $\mathcal{F}_{Linear}(\cdot)$, and a Sigmoid function $\mathcal{F}_{sig}(\cdot)$. The output of DCC module $\boldsymbol{F}^{*}\in \mathbb{R}^{C_{out}\times H \times W}$ is generated by using channel-wise calibration information to calibrate feature maps $\boldsymbol{F}_2$
\begin{equation}
\nonumber
    \boldsymbol{F}^{*} = \mathcal{F}_{sig}(\mathcal{F}_{Linear}(\mathcal{F}_{AvgPool}(\boldsymbol{F}_2)))\otimes \boldsymbol{F}_2.
\end{equation}

\subsection{Dynamically Calibrated Downsampling (DCD) module}
\begin{figure}
\centering
\includegraphics[width=0.9\textwidth]{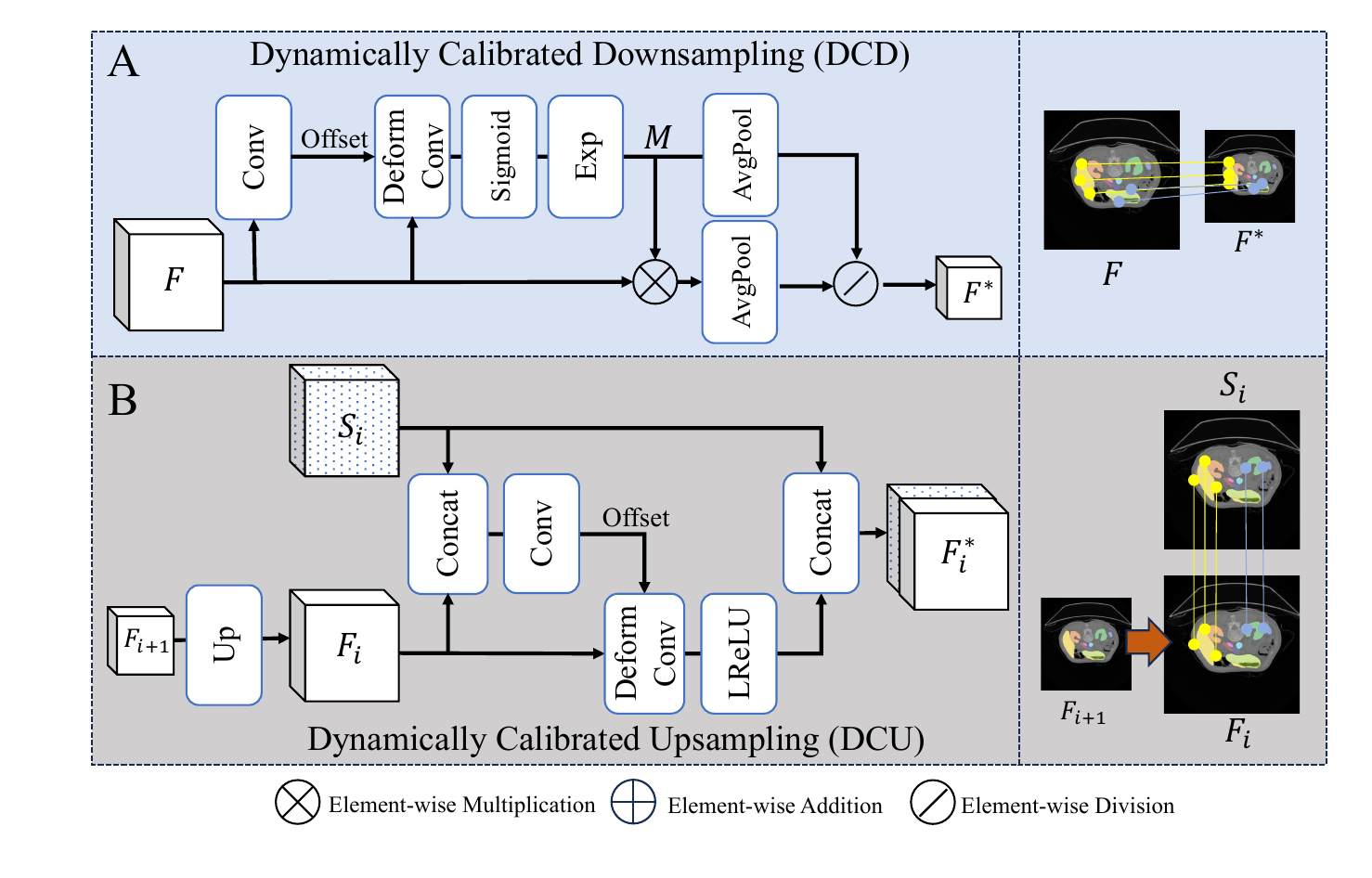}
\caption{The architecture of (A) the Dynamically Calibrated Downsampling (DCD) module and (B) the Dynamically Calibrated Upsampling (DCU) module.} \label{fig:fig2}
\end{figure}

The extensive use of downsampling layers (i.e., pooling and strided convolutions) prevents networks from preserving discriminative and deformable features. This limitation impedes the networks' ability to effectively model organ deformation. To tackle this limitation, we propose the Dynamically Calibrated Downsampling (DCD) module to adaptively downsample feature maps while preserving deformations and discriminative details (Figure~\ref{fig:fig2}(A)). First, the offset value $\boldsymbol{\Delta}_i\in \mathbb{R}^{C'\times H \times W}$ is learned from input features $\boldsymbol{F}=[\boldsymbol{f}_1,\boldsymbol{f}_2,...,\boldsymbol{f}_C]\in \mathbb{R}^{C\times H \times W}$ by a $3\times3$ convolution $\mathcal{F}_{conv}(\cdot)$ operation
\begin{equation}
\nonumber
    \boldsymbol{\Delta}_i  = \mathcal{F}_{conv}(\boldsymbol{F}).
\end{equation}
Subsequently, a deformable spatial feature allocation map $\boldsymbol{M}\in \mathbb{R}^{C\times H \times W}$ is learned by cascading a modulated deformable convolution $\mathcal{F}_{deform}(\cdot; \cdot)$, a Sigmoid function $\mathcal{F}_{sig}(\cdot)$, and an exponential operation $\mathcal{F}_{exp}(\cdot)$. This map describes how deformable spatial features can be allocated during downsampling and how input features contribute to final pooled features. The exponential operation $\mathcal{F}_{exp}(\cdot)$ is used to make the values in this map non-negative and easy to optimize.
\begin{equation}
\nonumber
    \boldsymbol{M} = \mathcal{F}_{exp}(\mathcal{F}_{sig}(\mathcal{F}_{deform}(\boldsymbol{F}; \boldsymbol{\Delta}_i)))
\end{equation}
Given the deformable spatial feature allocation map $\boldsymbol{M}$, the downsampled features $\boldsymbol{F}^*\in \mathbb{R}^{C\times \frac{H}{2} \times \frac{W}{2}}$ can be calculated. Specifically, features $\boldsymbol{F}$ are allocated by $\boldsymbol{M}$ as $\boldsymbol{M}\otimes \boldsymbol{F}$. Subsequently, allocated features are projected to a lower dimension by using the summation of allocation maps $\boldsymbol{M}$ to divide the summation of allocated features, where the summation is implemented by the average pooling $\mathcal{F}_{AvgPool}(\cdot)$.
\begin{equation}
\nonumber
    \boldsymbol{F}^* = \mathcal{F}_{AvgPool}( \boldsymbol{M}\otimes \boldsymbol{F})/\mathcal{F}_{AvgPool}(\boldsymbol{M})
\end{equation}
\subsection{Dynamically Calibrated Upsampling (DCU) module}
In hierarchical networks, feature maps upsampled in the decoding path are often fused with the skip-connected feature maps from the encoding path. However, Repeated downsampling and upsampling operations result in spatial misalignments between them. Thus, fusing them by either element-wise addition or concatenation will harm feature learning. To tackle this limitation, we propose the Dynamically Calibrated Upsampling (DCU) module to eliminate misalignments by adaptively aligning and calibrating features to its skip-connected reference before fusion (Figure~\ref{fig:fig2}(B)). First, the feature maps at the level $i$ in the decoding path $\boldsymbol{F}_i=[\boldsymbol{f}_1,\boldsymbol{f}_2,...,\boldsymbol{f}_C]\in \mathbb{R}^{C\times H \times W}$ were generated from those at the higher level $\boldsymbol{F}_{i+1}\in \mathbb{R}^{2C\times \frac{H}{2} \times \frac{W}{2}}$ by upsampling operations $\mathcal{F}_{up}(\cdot)$ (i.e., transposed convolutions)
\begin{equation}
\nonumber
    \boldsymbol{F}_i = \mathcal{F}_{up}(\boldsymbol{F}_{i+1}).
\end{equation}
Now, this upsampled feature maps $\boldsymbol{F}_i$ has the same dimension as its skip-connected reference $\boldsymbol{S}_i\in \mathbb{R}^{C\times H \times W}$ from the encoding path. Subsequently, feature maps (termed offset values $\boldsymbol{\Delta_i}\in \mathbb{R}^{C'\times H \times W}$ here) are extracted from both $\boldsymbol{F}_i$ and $\boldsymbol{S}_i$ by a $3\times3$ convolution $\mathcal{F}_{conv}(\cdot)$ operation. It is known that convolution operations can capture the spatial location information and present such information as feature maps. Thus, these offset values extracted from $\boldsymbol{F}_i$ and $\boldsymbol{S}_i$ can be considered as their spatial location information. When there are spatial misalignments between $\boldsymbol{F}_i$ and $\boldsymbol{S}_i$, these offset values demonstrate the shifted distances between each point in $\boldsymbol{F}_i$ and its corresponding point in $\boldsymbol{S}_i$
\begin{equation}
\nonumber
    \boldsymbol{\Delta}_i = \mathcal{F}_{conv}([\boldsymbol{F}_i, \boldsymbol{S}_i]).
\end{equation}
Given the learned offset $\boldsymbol{\Delta_i}$, a modulated deformable convolution $\mathcal{F}_{deform}(\cdot; \cdot)$ is applied to eliminate spatial misalignments. Specifically, this deformable convolution takes features $\boldsymbol{F}_i$ as input and adds these offset values $\boldsymbol{\Delta_i}$ to the regular locations of these features. Thus, the misalignments are eliminated by adjusting and aligning spatial locations of $\boldsymbol{F}_i$ to that of $\boldsymbol{S}_i$ using offset values within the deformable convolutional kernel. Finally, a Leaky ReLU $\mathcal{F}_{LReLU}(\cdot)$ is utilized to get the optimal feature maps $\boldsymbol{F}_i^*\in \mathbb{R}^{C\times H \times W}$, which can be concatenated with its reference $\boldsymbol{S}_i$
\begin{equation}
\nonumber
    \boldsymbol{F}_i^* = \mathcal{F}_{LReLU}(\mathcal{F}_{deform}(\boldsymbol{F}_i; \boldsymbol{\Delta}_i)).
\end{equation}

\section{Experiments}
\subsubsection{Datasets.}
We conducted experiments on two public datasets. The first one was the MICCAI 2021 FLARE Challenge dataset \cite{MA2022102616}. It consists of 361 abdominal CT images with voxel-level annotations of four organs. The second one was the MICCAI 2022 AMOS Challenge dataset \cite{ji2022amos}. It consists of 300 abdominal CT images with voxel-level annotations of 15 organs. Data were pre-processed and augmented by the nnUNet pipeline \cite{isensee2021nnu}.
\subsubsection{Experimental details.}
We used the combination of cross-entropy loss and Dice loss as the loss function and used the Dice Similarity Coefficient (DSC) to evaluate segmentation performance. Stochastic gradient descent (SGD) was used for network optimization. The initial learning rate was set to 0.01 and decayed using a poly decay strategy as nnUNet \cite{isensee2021nnu}. All experiments were performed using 5-fold cross-validation without additional post-processing.
\subsubsection{Baselines.}
We compared the segmentation performance between our Dynamic U-Net and U-Net baselines. The U-Net was implemented following the original design: max-pooling layers were used for downsampling and transposed convolution layers were used for upsampling. The SConv U-Net was implemented using strided convolution layers for downsampling, and other designs are the same with U-Net. Importantly, to demonstrate the effectiveness of each module, we conducted ablation studies by evaluating models with modules partially incorporated  (i.e., DCC U-Net, DCD U-Net, DCU U-Net). Specifically, for DCC U-Net, only the DCC module was incorporated into U-Net to replace convolution blocks in each layer. For DCD U-Net, only the DCD module was used for downsampling while the other aspects of the architecture remained the same as U-Net. For DCU U-Net, only the DCU module was used for upsampling while the other aspects of the architecture remained the same as U-Net.
\begin{table}
\centering
\caption{Comparison of multi-organ segmentation between Dynamic U-Net and baselines on the FLARE 2021. (\textbf{Bold} represents the best segmentation results).}\label{tab1}
\resizebox{1.0\textwidth}{!}{
\begin{tabular}{>{\centering\arraybackslash}p{2cm}||>{\centering\arraybackslash}p{1.5cm}|>
{\centering\arraybackslash}p{1.5cm}|>
{\centering\arraybackslash}p{1.5cm}|>
{\centering\arraybackslash}p{1.5cm}|>
{\centering\arraybackslash}p{1.5cm}|>{\centering\arraybackslash}p{1.5cm}}
\hline
Tasks &  U-Net$^\dagger$ &  SConv U-Net$^\dagger$ & DCC U-Net & DCD U-Net & DCU U-Net & Dynamic U-Net\\
\hline
Liver & 97.33 & 97.36 & $\boldsymbol{98.46}$ & 98.39 & 98.40 & 98.40 \\
Kidney & 95.50 & 95.58  & 96.66 & 96.70 & 96.62 & $\boldsymbol{97.22}$ \\
Spleen & 95.81 & 95.85 & 98.12 & 98.02 & 98.11 & $\boldsymbol{98.14}$\\
Pancreas & 79.91  & 80.66 & 81.83 & 81.38 & 81.23 & $\boldsymbol{83.54}$ \\
\hline
Average & 92.14 & 92.36 & 93.77 & 93.62 & 93.59  & $\boldsymbol{94.33}^*$ \\
\hline
\end{tabular}}\\
{\raggedright \scriptsize $^\dagger$: we implemented nnUNet with different architectures here.\\
$^*$: $p<0.01$ with Wilcoxon signed-rank test between Dynamic U-Net, and U-Net, SConv U-Net.
\par}
\end{table}
\subsubsection{Main Results.}
Table~\ref{tab1} and Table~\ref{tab2} show the result comparison of Dynamic U-Net with the baselines on multi-organ segmentation on the FLARE 2021 and AMOS 2022 benchmarks, respectively. The integration of the proposed DCC, DCD, and DCU modules in the proposed Dynamic U-Net resulted in a significant improvement in segmentation performance (Dice score) of both average foreground values and other organ-specific segmentation tasks compared with U-Net and SConv U-Net. To be specific, DCC U-Net achieved higher Dice scores across all tasks than U-Net, which demonstrates the effectiveness of DCC modules in the improvement of segmentation performance. This suggests that utilizing global contextual information may improve the representation capabilities of Dynamic U-Net. Furthermore, DCD U-Net and DCU U-Net achieved higher segmentation Dice scores than U-Net and SConv U-Net, demonstrating the improvement in the segmentation performance due to the proposed DCD and DCU modules. The DCD and DCU modules enable the Dynamic U-Net to adaptively preserve and calibrate deformable and discriminative features, thereby effectively addressing large variations among different organs.
\begin{table}
\centering
\caption{Comparison of multi-organ segmentation between Dynamic U-Net and baselines on the AMOS 2022. (\textbf{Bold} represents the best segmentation results).}\label{tab2}
\resizebox{1.0\textwidth}{!}{
\begin{tabular}{>{\centering\arraybackslash}p{2.5cm}||>{\centering\arraybackslash}p{1.4cm}|>
{\centering\arraybackslash}p{1.4cm}|>
{\centering\arraybackslash}p{1.4cm}|>
{\centering\arraybackslash}p{1.4cm}|>
{\centering\arraybackslash}p{1.4cm}|>{\centering\arraybackslash}p{1.5cm}}
\hline
Tasks &  U-Net$^\dagger$ &  SConv U-Net$^\dagger$ & DCC U-Net & DCD U-Net & DCU U-Net & Dynamic U-Net\\
\hline
Spleen & 96.31&96.60 & 96.83 & 97.05 & 97.12& $\boldsymbol{97.64}$  \\
R. kidney & 96.45&96.15  & 96.61 & 97.18 & 97.03 &  $\boldsymbol{97.70}$\\
L. kidney & 96.36&96.23  & 96.68 & 96.81& $\boldsymbol{96.99}$ & 96.33 \\
Gall bladder & 78.58&79.56  & 79.77 & 80.50& 80.22 & $\boldsymbol{83.83}$ \\
Esophagus & 84.13&83.61  & 84.84 &85.37& 85.25 & $\boldsymbol{86.17}$ \\
Liver & 96.74&96.76  & 96.97 & 97.53& 97.36 &  $\boldsymbol{98.52}$ \\
Stomach & 87.39&88.65  & 89.44 & 89.32& 89.62 & $\boldsymbol{90.29}$ \\
Arota & 94.54&94.46  & 94.71 & 95.24& 95.27 & $\boldsymbol{96.65}$ \\
Postcava & 89.02&88.78 & 89.90 & 90.25& 90.06 & $\boldsymbol{90.36}$  \\
Pancreas & 83.93&84.50  & 85.52 & 85.67 & 86.38& $\boldsymbol{87.06}$  \\
R. adrenal gland & 77.71& 78.27 & 78.74 & 78.65& 79.50 & $\boldsymbol{80.47}$ \\
L. adrenal gland & 79.65 &79.58 & 81.29 & 81.32 & 81.16 & $\boldsymbol{81.44}$ \\
Duodenum & 77.49 &77.87 & 79.04 & 78.79& 80.22 &  $\boldsymbol{81.11}$ \\
Bladder & 86.40&85.40 & 88.12 & 89.04& 88.90& $\boldsymbol{89.06}$ \\
Prostate/uterus & 80.56 & 80.97 & 83.30 & $\boldsymbol{84.52}$ & 84.49 & 83.81 \\
\hline
Average & 87.02 & 87.16  & 88.12 & 88.48 & 88.64 & $\boldsymbol{89.36}^*$ \\
\hline
\end{tabular}}
{\raggedright \scriptsize $^\dagger$: we implemented nnUNet with different architectures here.\\
$^*$: $p<0.01$ with Wilcoxon signed-rank test between Dynamic U-Net, and U-Net, SConv U-Net.
\par}
\end{table}
\section{Conclusion}
We introduce a novel Dynamic U-Net for abdominal multi-organ segmentation from CT images, and it is proposed to handle the morphological complexity and large variation among different organs by incorporating Dynamically Calibrated Convolution, Dynamically Calibrated Downsampling, and Dynamically Calibrated Upsampling modules. It was empirically shown that the incorporation of these modules may continuously improve segmentation accuracy.

\begin{figure}
\centering
    \centering
    \includegraphics[width=\textwidth]{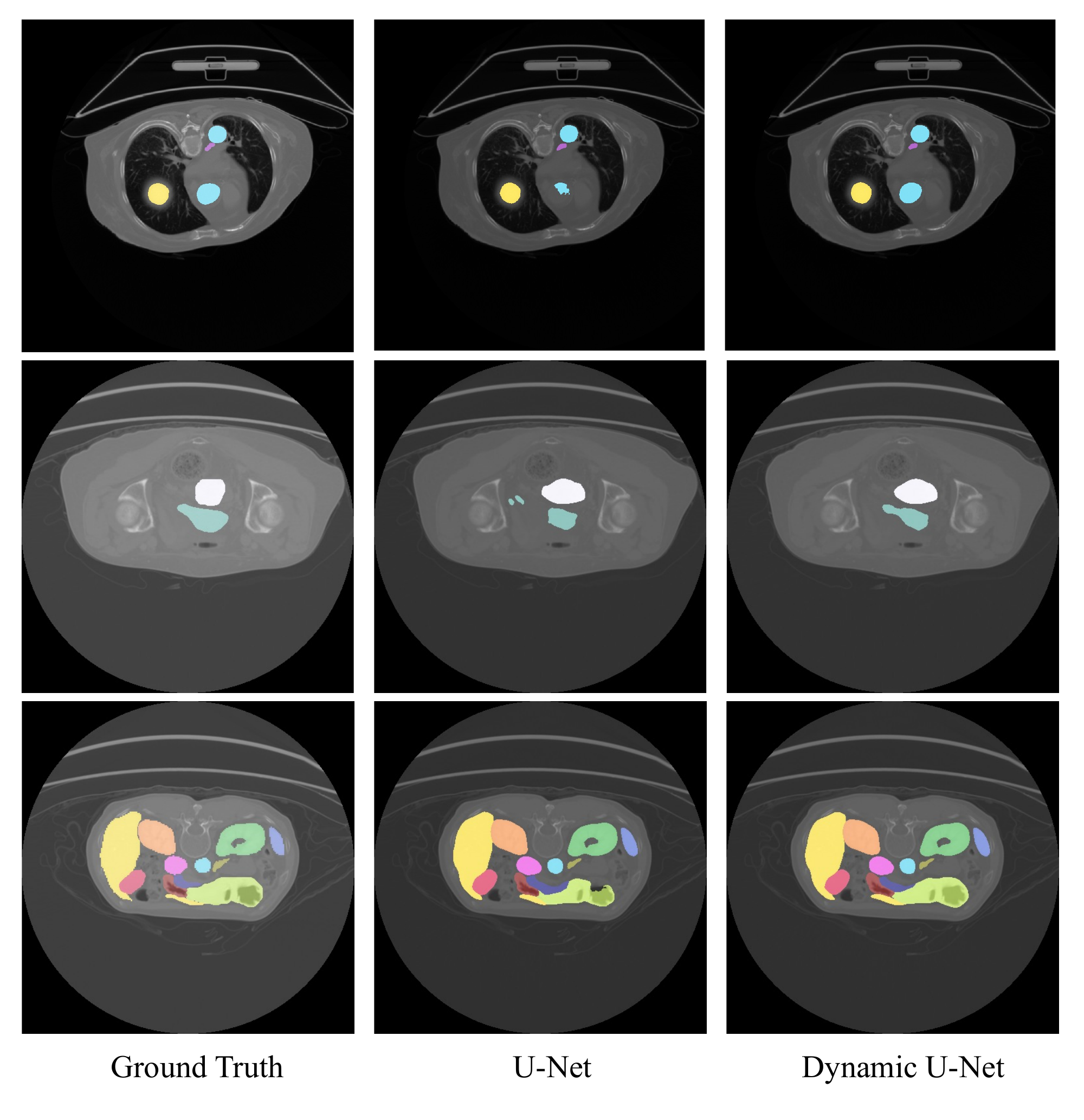}
    \caption{Qualitative representations of multi-organ segmentation on AMOS 2022 dataset. Dynamic U-Net shows better segmentation quality than U-Net.} 
\end{figure}

\subsubsection{Acknowledgements} This research was funded by 5 U24 CA258483. Computations were performed using the facilities of the Washington University Center for High Performance Computing (CHPC), which was partially funded by National Institutes of Health (NIH) grants S10OD025200, 1S10RR022984-01A1, and 1S10OD018091-01.

%
%
%
\bibliographystyle{splncs04}
\bibliography{DynamicUNet}

\end{document}